# Implementation of Design Changes Towards a More Reliable, Hands-off Magnetron Ion Source

A. Sosa[1, a)], D.S. Bollinger[1], P.R. Karns[1] and C.Y. Tan[1]

[1]*Fermi National Accelerator Laboratory, P.O. Box 500, Batavia, IL 60510-5011, USA*

a)Corresponding author: asosa@fnal.gov

**Abstract.** As the main H- ion source for the accelerator complex, magnetron ion sources have been used at Fermilab since the 1970's. At the offline test stand, new R&D is carried out to develop and upgrade the present magnetron-type sources of H⁻ ions of up to 80 mA and 35 keV beam energy in the context of the Proton Improvement Plan. The aim of this plan is to provide high-power proton beams for the experiments at FNAL. In order to reduce the amount of tuning and monitoring of these ion sources, a new electronic system consisting of a current-regulated arc discharge modulator allow the ion source to run at a constant arc current for improved beam output and operation. A solenoid-type gas valve feeds $H_2$ gas into the source precisely and independently of ambient temperature. This summary will cover several studies and design changes that have been tested and will eventually be implemented on the operational magnetron sources at Fermilab. Innovative results for this type of ion source include cathode geometries, solenoid gas valves, current controlled arc pulser, cesium boiler redesign, gas mixtures of hydrogen and nitrogen, and duty factor reduction, with the aim to improve source lifetime, stability, and reducing the amount of tuning needed. In this summary, I will highlight the advances made in ion sources at Fermilab and will outline the directions of the continuing R&D effort.

## INTRODUCTION

C.W. Schmidt developed a version of the magnetron ion source at Fermi National Accelerator Laboratory (FNAL) in the late 1970's. J. Alessi at Brookhaven National Laboratory (BNL) further optimized the FNAL design, reducing the discharge current, increasing the extraction voltage and introducing a dimpled cathode that used permanent magnets [1]. Operational experience from BNL has shown that this type of source is more reliable with a longer lifetime due to better power efficiency [2]. The magnetron source design currently used at FNAL, is similar to both the BNL ion source and the source developed by C.W. Schmidt for the High Intensity Neutrino Source (HINS) project at FNAL [3]. The source has a round aperture with direct extraction using a 45° extraction cone. The source cathode has a spherical dimple which provides focusing of the H- ions leaving the surface to the anode aperture increasing the power efficiency to 48 mA/kW. The source is mounted reentrant inside a 10-inch stainless steel cube and was designed with "ease of maintenance" in mind. The extraction method is single stage with the extraction cone at ground potential and the source biased to -35 kV. This bias voltage is pulsed at 15 Hz. This paper summarizes the studies and modifications done in the source over the last three years with the aim of improving its stability, reliability and overall performance.

## SOLENOID GAS VALVE

The magnetron H$^-$ ion sources currently in operation at Fermilab use Veeco PV-10 piezoelectric gas valves to pulse H$_2$ gas into the ion source [4]. A Viton disk has been glued to the center of the valve and provides a vacuum seal. In general, these piezoelectric valves work well and are very reliable for our application but their flexibility tends to decrease with increasing temperature. As room temperature increases, the piezoelectric valve flexes less, allowing less gas into the source, thus affecting the arc current and beam output. A change of 1 °C in room temperature can translate into a 1 µTorr change in vacuum pressure and as much as 1 A change in arc current. Operators frequently need to adjust gas valve voltage and/or gas pulse width to keep up with changes in room temperature. In addition, calibration issues with the installation of the gas valve and its tension spring account for significant performance differences from one piezoelectric valve to the next. This inconvenience motivated the need to find an alternative way of injecting H$_2$ gas into the source to avoid these issues. A commercial pulsed solenoid valve has been characterized in a dedicated off-line test stand to assess the feasibility of its use in the operational ion sources. The valve is a Series 9 by Parker Hannifin Corp. The performance of the solenoid gas valve has been characterized by measuring the beam current output of the magnetron source with respect to the voltage and pulse width of the signal applied to the gas valve. A VESPEL® poppet was used due to its thermal and mechanical properties [5]. The valve aperture is 0.5 mm in diameter. H$^-$ ion beams with beam currents >50 mA have been extracted at 35 keV using the solenoid gas valve.

## CESIUM OVEN

The magnetron ion source is a Surface Plasma Source (SPS) [6], which relies on cesium to lower the work function of the cathode for H$^-$ production [7]. A cesium oven is used to heat elemental cesium so that it flows to the rest of the cesium transport system to replenish the cesium layer on the cathode, which is depleted due to the high cathode temperature and plasma erosion. The cesium oven is used to cesiate the cathode in the ion source. The cesium ovens currently in operation at FNAL are 139.7 mm long, 14.29 mm diameter tubes made of copper (Fig. 1a). A commercially available 5 g cesium vial is inserted into the copper tube, then sealed on top with an isolation valve, pumped down and then the oven valve is closed and the copper tube is pinched with a tool to crack the cesium vial inside and release it. This method has worked well for many years, but it presents several challenges. Glass fragments from the cracked cesium vial pose a risk when performing cleaning and maintenance tasks on used cesium ovens. Another issue with the cesium oven is that, copper being a very good heat conductor made it hard to obtain the low nominal oven temperature of 110°C as the cesium oven is heated not only by the heat tape, but by the valve heater as well. The heat tape had to be run at very small currents along with thinning out the exterior cesium oven insulation. As a result, the cesium oven temperature would start following the room temperature. A new cesium oven made of stainless steel was designed to fix these problems (Fig. 1b). The more compact design helps with thermal insulation of the cesium oven from the rest of the cesium delivery systems. The smaller design also requires pouring the cesium into it from the glass vials.

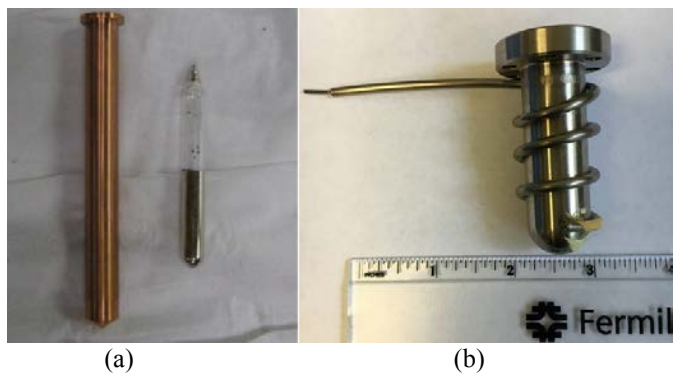

(a)          (b)

**FIG. 1.** a) Cesium oven made of copper (left) and vial with 5 g of elemental cesium (right). b) New cesium oven made of stainless steel with 2 Ω/m heating wire around it.

## CURRENT-REGULATED ARC MODULATOR

The arc pulser currently used in the operational sources is voltage regulated. The main inconvenience of running an ion source with a voltage-regulated arc modulator is the drift of the arc current over time. Since arc current and beam output are intimately related, a drifting arc current means an unstable beam output over the course of beam operations. In order to avoid this, a current-regulated arc modulator was designed at FNAL, inspired by the current-regulated arc modulator from BNL. It consists of a high-voltage card and a trigger/feedback loop card. The voltage to current converter comprises an op-amp set up as a positive high gain amplifier connected to a power MOSFET which acts as a gate for controlling the large current flowing out of the high voltage power supply. The high current power MOSFET IXFN132N50P3 [8] is spark resistant at -35 kV. It has a forward DC current rating of 112 A and a drain-source voltage rating of 500 V. The fast rise and fall times of ~100 ns of this MOSFET help significantly reduce the fall and rise times of the arc current pulse.

## BEAM CURRENT NOISE

An experiment was done in the offline test stand to mix $H_2$ and $N_2$ gases very accurately at different ratios on the gas line that feeds the source with the aim of exploring the effect of different $N_2/H_2$ ratios on the beam current noise. $N_2/H_2$ ratios between 0.5 to 3% could be tested, and the mean beam current noise was measured connecting the output of the current toroid to an oscilloscope. For every data point, the maximum beam current was recorded at an extraction voltage of -35 kV. Noise ratio slightly increases as more nitrogen is added to the source, however the noise ratio remains low, at under 1%. The observed trend in the raw data also reflects that, as more nitrogen is added to the source, the beam current signal decreases while the standard deviation of the signal remains unchanged, hence the increase in the noise ratio. Above 3% $N_2$ the arc current is severely affected or even terminated in a matter of minutes. Noise ratio was also studied for different arc currents at the same extraction voltage of -35 kV. The extracted beam current increases linearly with arc current, therefore a lower noise ratio is observed at higher arc currents with and without $N_2$. Looking at raw data, the only magnitude that changes appreciably is the beam current signal but not its standard deviation. The addition of nitrogen to the ion source in the conditions of these experiments was not useful in reducing the beam current noise.

## CONCLUSIONS

The magnetron ion source can deliver beam currents above 50 mA with the solenoid valve, which satisfies the needs of laboratory. The new cesium oven provides the ability to fine tune the temperature with little current increments. The new arc modulator is a major improvement towards more stable beam output for operations. No reduction in the noise of the beam current or arc discharge was observed with the proposed ideas. The magnetron ion source is inherently noisy and the noiseless discharge regime of the magnetron seems to be in an unprofitable region to deliver beam operationally.

## ACKNOWLEDGMENTS

Operated by Fermi Research Alliance, LLC under Contract No. DE-AC02-07CH11359 with the United States Department of Energy. The authors would like to thank A. Feld and K. Koch for their technical support with the gas valves and the ion source during these experiments. I also want to thank T. Lehn (BNL) for sharing the design notes of their current regulated arc modulator.